\documentclass[doublecol]{epl2}

\revision

\title{Criticality and Continuity of Explosive Site Percolation in Random Networks}
\shorttitle{Criticality and Continuity of Explosive Site Percolation in Random Network}

\author{ J. H. Qian\inst{1} 
\and D. D. Han\inst{2}\thanks{E-mail: \email{ddhan@ee.ecnu.edu.cn}}
\and Y. G. Ma\inst{1}\thanks{E-mail: \email{ygma@sinap.ac.cn} }
}
\shortauthor{J. H. Qian \etal}

\institute{
  \inst{1} Shanghai Institute of Applied Physics, Chinese Academy of Sciences,
Shanghai 201800, China\\
  \inst{2}  School of Information Science and
Technology, East China Normal University, Shanghai 200241, China}

\pacs{89.75.Hc}{Networks and genealogical trees}
\pacs{89.75.Da}{Systems obeying scaling laws}

\abstract{This Letter studies the critical point as well as the
discontinuity of a class of explosive site percolation in Erd\"{o}s
and R\'{e}nyi (ER) random network. The class of the percolation is
implemented by introducing a \textit{best-of-m} rule. Two major
results are found: i). For any specific $m$, the critical
percolation point scales with the average degree of the network
while its exponent associated with $m$ is bounded by $-1$ and
$\sim-0.5$. ii). Discontinuous percolation could occur on sparse
networks if and only if $m$ approaches infinite. These results not
only generalize some conclusions of ordinary percolation but also
provide new insights to the network robustness.}

\begin{document}

\maketitle

  Percolation describes the connectivity of a graph by continuously
occupying links or nodes. Due to its wide application in a variety
of area such as epidemics, nuclear multifragmentation and network
robustness etc., it has been an active subject of research for
decades\cite{new,shu,dorovge,nuclei,Hal,new2}. Perhaps of the
greatest importance in percolation studies is its order of phase
transition and the location of critical point. In almost all cases
percolation on graphs is shown to be continuous (higher than first
order) while critical points for networks of various structure and
dimension have been determined. In an ER random network with average
degree $k$, as a well studied case, ordinary site percolation causes
a continuous phase transition at critical point
$t_c(k)=1/k$ \cite{stan,rozen}.

 Recently a new kind of percolation, named explosive percolation was
proposed. By introducing a proper competitive mechanism, it was
first found by Achlioptas, D'Souza, and Spencer and was
subsequently studied intensely by other scientists that the bond
percolation in random networks could be discontinuous
\cite{Ach,ziff,Kim,Fortu,Frie,localagg,Hermann,tricri,chen}.
However,  further numerical and theoretical studies demonstrated
that such percolation is actually continuous in the thermodynamic
limit but has unusually small critical exponent
\cite{dorovge2,natureph,peter,rio}. But this is not the end of the
story. For instance, a very recent study pointed out that the
behavior of the explosive  percolation transition depends on
detailed dynamic rule \cite{cho2}. When dynamic rules are designed
to suppress the growth of all clusters,  the explosive percolation
transition could be discontinuous. Another recent study on
explosive site percolation on square lattice also claimed  the
existence  of discontinuous phase ]transition \cite{woo}.
Therefore whether the explosive  percolation  is indeed
discontinuous or continuous is still controversial.

Despite the extensive studies on discontinuity of explosive
percolation, the other important property, namely the location of
the critical point, has not been studied systematically. Previous
studies presented the related results only for their special models,
which neither provide any general conclusions nor help to understand
its physical meaning in network dynamics. In this Letter, critical
percolation point is the prior issue to discuss. This is partly
motivated by some earlier works on network robustness under attack
\cite{bara,Hal2}. While most of their attack strategies are based on
degree information, real-world networks may suffer more diverse
structure frangibility to which a practical attack strategy could be
more specific. The concept of explosive percolation allows us to
explore these questions in a general way and could provide new
insights to network robustness. One of the aim of the Letter is to
show that in addition to discontinuity or not, explosive percolation
has other appealing properties worth studying. The discontinuity of
the percolation and the related critical exponents are also
discussed properly.

\begin{figure} \resizebox{16pc}{!}{\includegraphics{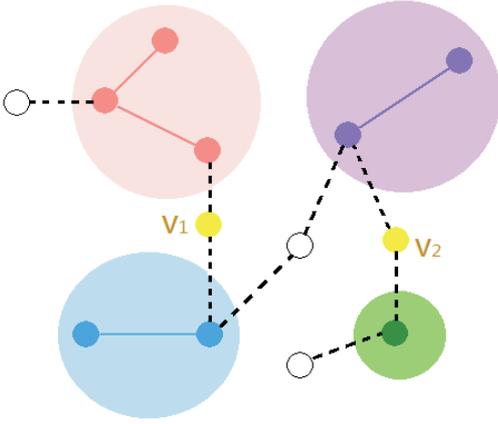}}
\vspace{-.5cm}
 \caption{\footnotesize Visualization
of the explosive rule for the case of $m=2$. The empty nodes are
unoccupied nodes. Four circles of different color and the
corresponding inside nodes represent four occupied clusters. Two
golden nodes ($V_1$ and $V_2$) are the candidates selected randomly
and compete for occupation. The one connecting the clusters of the
smallest total size will be chosen. In this case, the node $V_2$ is
finally chosen since it connects the clusters of total size $3$,
smaller than $5$, the total size of the clusters that $V_1$
connects. Note that the candidates are selected from all possible
empty nodes. If a selected candidate is an isolated single node of
ER network, this node itself represents a cluster of size equaling
to $1$. } \label{model}
\end{figure}

\begin{figure} \resizebox{20pc}{!}{\includegraphics{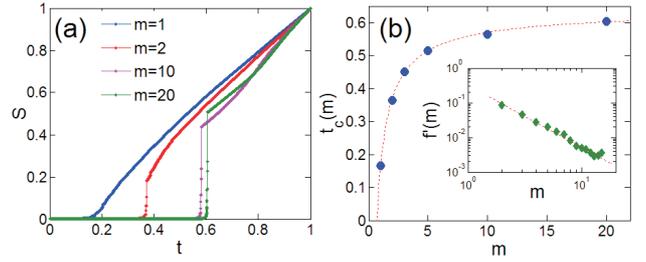}}
\vspace{-.3cm} \caption{\footnotesize   (a): Site percolation
process for the case of $k=6$. With the increase of $m$, critical
point becomes much more delayed, but its delayed rate decreases
rapidly. Eg., the delayed time is $0.198$ between $m=1$ and 2 but
only $0.038$ between $m=10$ and $20$. The change of $m$ is ten times
larger but the delayed rate is five times smaller than the previous
one. The figure is a single simulation for network size of $5\times
10^{5}$ nodes. (b): The increase of $t_c$ with $m$ for the case of
$k=6$. Blue solid point is the simulated data while red dashed line
is the fit by using Eq.(2). In this case, $T_c=0.633,
f(m)=-0.490m^{-0.794}$. Inset: the power-law decay of $f'(m)$
obtained by calculating the difference of $t_c(m)$ for successive
$m$ from $1$ to $16$. In this simulation, numerical calculations for
$t_c(m,k,N)$ are carried out up to network size of $10^{5}$ nodes
and are averaged over 50 realizations.} \label{transition}
\end{figure}

Let us begin with our explosive rule. Consider a connected ER random
network with total number of nodes $N$ and average degree $k$. At
each time step $t$, defined as the fraction of number of occupied
nodes of $N$, $m$ empty nodes are selected randomly as candidates to
be occupied, but only the one which minimizes the sum of the size of
the clusters that itself connects is finally chosen, as visualized
in Fig~\ref{model}. If there is more than such a node, we choose one
of them randomly. This competitive process is the so-called
\textit{best-of-m} rule which has also been applied to bond
percolation \cite{tricri,natureph}. Repeating this process, the
system will eventually become percolated at a critical point
$t_c=t_c(m,k,N)$, while its critical behavior depends crucially on
the value of $m$. With the increase of $m$, the growth of the order
parameter $S$, defined as $S=S_1/N$ where $S_1$ is the the size of
the largest connected cluster, is effectively suppressed, leading to
gradually delayed critical points and a discontinuous-like jump in
its critical dynamics (Fig.~\ref{transition}(a)). For the case of
$m=1$, the traditional continuous site percolation is recovered
while for $m=N$ the system becomes the most explosive. In most
cases, percolation concerns the critical behavior in the
thermodynamics limit, so we focus on critical point in the
thermodynamics limit $t_c(m,k)$, which is defined as
$t_c(m,k)=t_c(m,k,N\rightarrow\infty)$.

Further inspection on Fig.~\ref{transition}(a) reveals that the
increase rate of the critical point decreases with $m$. In fact, the
increase rate vanishes so quickly that the critical point is
conjectured to have a nontrivial limitation
$T_c(k)=t_c(m=N\rightarrow\infty,k)<1$ . This conjecture can be
reached by contradiction. Suppose such limitation does not exist,
i.e. $T_c=1$, it is expected that just before this $T_c$, the
network has only $N^{\eta}$ $(\eta<1)$ unoccupied nodes and
$N^{1-\alpha}$$(\alpha<1)$ isolated occupied clusters of similar
size of $O(N^{\alpha})$ as global competition in explosive bond
percolation.\cite{natureph}. Denoting $k'$ the average degree within
each of the cluster, these clusters surely belong to the ensemble of
subgraph of $N^{\alpha}$ nodes and $k'N^{\alpha}/2$ links. On the
other hand in ER random network, the average number of appearance
$G$ of such subgraph  can be calculated as 
\begin{equation}
\langle G \rangle \sim N^{(1-\alpha)N^{\alpha}(1-\frac{k'}{2})}.
\end{equation}
Consistently we should have $\langle G \rangle \geq N^{1-\alpha}$,
leading to $k'\leq 2$. However $k'\geq 2$ must hold because each
occupied cluster is connected. Then in the thermodynamic limit it
comes to the only possible solution $k'\rightarrow 2$. While for a
connected random graph, the average degree $k$ is generally larger
than $2$, so for each of the isolated cluster there are about
$(k-2)N^{\alpha}$ links stem from the cluster to the remaining
unoccupied nodes, leading to $O(N)$ such links totally. We thus
arrive at a contradiction since the unoccupied nodes can only
provide $kN^{\eta}/2<O(N)$ links. Note that although the proof is
made for $k>2$, it does not rule out the possibility of its
validation for $k<2$. Actually $T_c=1$ only occurs when
$k\rightarrow 1$ as will be discussed later.

\begin{figure} \resizebox{20pc}{!}{\includegraphics{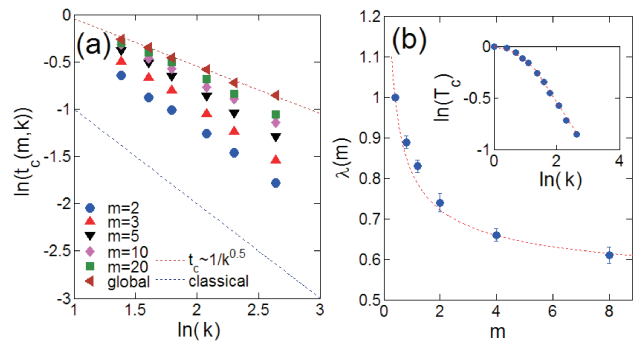}}
\vspace{-.3cm}
 \caption{\footnotesize   (a):
Correlation of $t_c(m,k)$ with $k$ for different $m$. All of them
follow power-law correlation with exponent decreasing with $m$,
ranging from $\tau\approx0.5$ to $1$. Blue line represents the
result of ordinary percolation, i.e. $t_c(1,k)=1/k$. Brown  line
represents Eq.(3). $T_c(k)$ (brown left-triangle) for
$k={4,5,6,8,10,14}$ is 0.757(4), 0.682(2), 0.633(3), 0.554(5),
0.484(3), 0.421(3), respectively, where the number in the bracket
refers to the error of the last digit. Eg. 0.757(4) means 0.757
$\pm$ 0.004, respectively. Note that this error is a
total error, i.e. it includes errors from a finite-size scaling of $t_c(m)$ and Eq.(2)} (b): Correlation of the
exponent $\lambda(m)$ with $m$. Blue solid point is the simulation
results while red line represents Eq.(5). Inset: Full correlation of
$T_c$ with $k$. The prediction by Eq.(6) (red line) coincides well
with the simulation (blue solid point). In both simulations,
numerical calculations for $t_c(m,k,N)$ is carried out up to network
size of $10^{5}$ nodes and is averaged over 50 realizations.
\label{tck}
\end{figure}

It is then nontrivial to find such limitation $T_c$. Defining the
susceptibility as $\chi\equiv N\sqrt{\langle S^2 \rangle- \langle
S \rangle^2}$, which quantifies the amplitude of the fluctuations
of the size of the largest cluster, the critical point
$t_c(m,k,N)$ of finite network size $N$ is said to locate at which
$\chi$ reaches its maximum. Then with a finite size scaling
function $t_c(m,k,N)-t_c(m,k)\sim N^{-1/\nu}$, the critical point
in the thermodynamic limit $t_c(m,k)$ can be precisely determined
\cite{Fortu}. By this method, $t_c(m,k)$ for each
$m={2,3,5,10,20}$ and each $k={4,5,6,8,10,14}$ is calculated as a
preparation for finding $T_c$. $T_c$ for a particular $k$ can be
written as
\begin{equation}
t_c(m)=T_c+f(m).
\end{equation}
It is valid that $t_c(m+1)-t_c(m)=f(m+1)-f(m)\approx f'(m)$ , which
indicates that the formula of $f(m)$ can be derived by calculating
the difference of $t_c(m)$. Then further calculation of $t_c(m)$ for
successive $m$ from $1$ to $16$ is made and $f(m)$ is found to
follow a power-law formula (inset of Fig.~\ref{transition}(b)). Having
known the expression of $f(m)$, Eq.(2) can be used to determine
$T_c$ (Fig.~\ref{transition}(b)). The major error of this method
results from the finite number of the fitted data points. The error
can be evaluated in the following way. Consider that we have a
sequence of data points of successive $m$ up to $m=m_f$, the fit for
these data points gives $T_c^{m_f}$. Then the error is given by
$|T_c^{m_f}-T_c| $= $|T_c^{m_f}-T_c^{m_f+1}+T_c^{m_f+1}$
$-T_c^{m_f+2}+...+T_c^{\infty}-T_c| $ $\leq \int^{\infty}_{m_f}|T_c^{x}-T_c^{x+1}|dx$.
We calculate $T_c^{m_f}$ for successive $m_f$ and find that
$|T_c^{m_f}-T_c^{m_f+1}|$ follows a power-law formula. In the case
of Fig.~\ref{transition}(b), for example, the formula approximately
equals $0.8m_{f}^{-3}$. With $m_f=20$, the error is estimated to be
smaller than $10^{-3}$. Actually our calculation for each $T_c(k)$
indicates that the error caused by the proposed finite size scaling
method is no more than $O(10^{-3})$. Detailed results of $T_c(k)$ are
reported in the legend of Fig.~\ref{tck}.

To find a general correlation of critical point and the network's
structure, $T_c$ is plotted as a function of the average degree
$k$(Fig.~\ref{tck}). It gives
\begin{equation}
T_c(k)\sim k^{-\tau},
\end{equation}
where $\tau=0.49\pm0.02$. Rather interestingly, $T_c(k)$ still
scales as a power law but with a different exponent $-\tau\approx
-0.5$, in contrast to $-1$ found in ordinary site percolation. What
is more nontrivial is that a general expression of $t_c(m,k)$ can be
directly deduced from Eq.(3). Since for all $1<m<N$, $t_c(m,k)$ is
bounded by two power-law functions, i.e. Eq.(3) and $t_c(m=1,k)=1/k$
(the critical point of ordinary percolation), according to the
argument in Ref.\cite{Han} critical point $t_c(m,k)$ for all $m$
must scale as a power law with $k$ and its exponent depends on the
specific value of $m$.
\begin{equation}
t_c(m,k)\sim k^{-\lambda(m)}.
\end{equation}
Fig.~\ref{tck} demonstrates the validation of Eq.(4). The exponent
$\lambda(m)$ is found to be well approximated by an
\textit{arctan} function
\begin{equation}
\lambda(m)\approx\frac{\pi}{4}\frac{1}{\arctan{m^{0.4}}} .
\end{equation}
Although the validation of Eq.(5) needs to be further examined since
it is a conjectured result, it coincides with $\lambda(1)=1$,
$\lambda(m=N\rightarrow\infty)=0.5$ and also fit the body well as
shown in Fig.~\ref{tck}(b).

What differs from the ordinary percolation is that the scaling
(Eq.(4)) is not always valid but deviates from power law for small
$k$ when $m>1$, as demonstrated in the inset of Fig.~\ref{tck}(b)
for the extreme case of $m=N$ (i.e. $T_c$). Actually the deviation
becomes more apparent as $m$ grows. This deviation from the power
law origins from two aspects: the unconnectedness of the underlying
ER random graph for small $k$ and the preferential selection for
isolated nodes in explosive percolation. When $k=2$, for example,
despite of the existence of the giant component of the underlying
graph there are still about $20$ percent of nodes forming isolated
small clusters. In contrast to an absolutely uniform occupation in
ordinary percolation, these isolated nodes are preferentially
occupied before criticality in the explosive ones but have no
contribution to the emergence of the giant percolated cluster.
However, in a connected graph every node helps to bring out giant
cluster during the process of percolation. Therefore the constituent
of $t_c(m>1,k)$ of small $k$ includes many \textit{idle} nodes and
thus causes the deviation of power law. If $m$ grows, the degree of
preferential occupation of isolated nodes increases, so the
deviation becomes more apparent. To make the explanation more
convincing, $T_c$ for small $k$ will be reconstructed according to
the above explanation. In the case of
$m=N\rightarrow\infty$, it is expected that $i)$ all the underlying
isolated nodes will be occupied before criticality; $ii)$ the whole
network will not be percolated until the underlying giant component
is percolated. Therefore $T_c$ can be written as the sum of two
parts: the relative size of those isolated nodes and the relative
size of the occupied nodes at criticality within the giant component
of the underlying graph, as expressed in the following equation
\begin{equation}
T_c(k)=1-S_r+T_c(\kappa(k))S_r,
\end{equation}
where $S_r$ is the relative size of the giant component of the
underlying graph which satisfies the well-known equation
$S_r+e^{-kS_r}=1$ \cite{bolla}. $T_c(\kappa(k))$ is the
critical point of the percolation on the underlying giant component
whose average degree is $\kappa(k)$. Clearly $1-S_r$ is
the relative size of the isolated nodes while $T_c(\kappa(k))S_r$
represents the contribution of the occupied nodes within the
underlying giant component.  Since $S_r\rightarrow0$ as
$k\rightarrow1$, it is clear that the critical point occurs at $1$
for $k=1$. For $k>1$ the prediction of $T_c$ depends on $\kappa(k)$
which is found numerically as
$\kappa(k)=2.337e^{-0.855k}+k$. Therefore to obtain
$T_c(k)$, we need to know $T_c(\kappa(k))$, and consequently
$T_c(\kappa_{2}(k))$, where $\kappa_{2}(k)=\kappa(\kappa(k))$, and
so on. Finally, we will arrive at a certain $l$ so that
$\kappa_{l}(k)>4$. Having known the validation of Eq.(3) for $k>4$,
the above-mentioned method allows a recursive estimation for $T_c$.
As shown in the inset of Fig.~\ref{tck}(b), the prediction coincides
the simulation results well, which confirms our explanation.

Eq.(3) to Eq.(5), as the main contribution in our study, not only
generalize some classical conclusions but could also give some new
insights to the network robustness under attack. While most related
studies designed their attack strategy based on some topology
properties such as degree \cite{bara,Hal2}, a practical attack
strategy depends totally on how much the attacker knows about the
network. If the attacker gets no information on network, random
attack is the only option (the case of $m=1$). But if the attacker
knows the network structure every detail, he/she can find the best
way (not just based on degree or some other properties) to destroy
the network (the case of $m=N$). Thus the practical meaning of $m$
could be a measurement of the information that the attacker has on
the network. Our study indicates that it is possible to understand
the network robustness to any kind of attack based on any such
information. However, since assessments of these information are
usually impractical, Eq.(3) shows its great significance because
$T_c$ can be served as a warning that possible breakdown could ever
occur once the attack process exceeds this point. These concepts and
ideas are nontrivial in the sense that they indicate a possible
general framework for the study of network robustness. Indeed from
the viewpoint of information, the network robustness might be
defined as a strategy-independent function, namely $t_c(m)$ while
previous strategy-dependent work can hardly provide such general
definition due to their unquantifiable model.

\begin{figure} \resizebox{20.5pc}{!}{\includegraphics{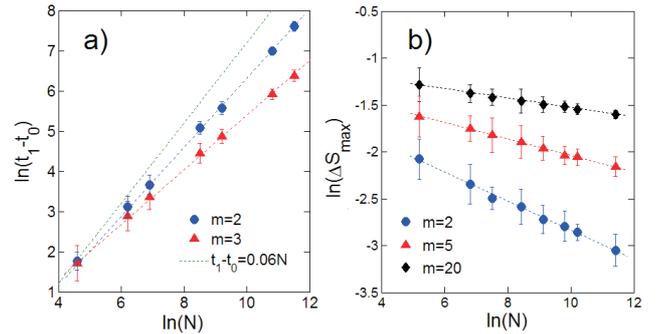}}
\vspace{-.3cm} \caption{\footnotesize   (a): The
scaling of $\Delta t$ with $N$. For $m=2$, we find
$\alpha\approx0.5$ and $c\approx0.15$, leading to $\Delta t\sim
N^{0.84}$ (pink line). For $m=3$, we find $\alpha\approx0.6$ and
$c\approx0.25$, leading to $\Delta t\sim N^{0.66}$ (black line).
Both of the interval ${\Delta t}/{N}$ vanish in the thermodynamics
limit. But for the traditional percolation, the interval always
extends, i.e. $\Delta t\sim O(N)$, as indicated by green line. (b):
The power-law relationship of $\Delta S_{max}$(the size of the
largest jump of the order parameter) and network size $N$. The
exponents for the case of $m=2,5,20$ are measured as
$0.15,0.088,0.05$, respectively. All the results are obtained under
network size up to $10^{5}$ nodes and are averaged over $50$
realizations.} \label{explo}
\end{figure}

Now let us turn to the discontinuity of the explosive site
percolation. The explosive nature can be established in the
following way \cite{Ach}. Let $t_0$ denotes the last step for which
$S_1<N^\alpha$, and $t_1$ the first step for which $S_1>cN$, where
$\alpha$ and $c$ are two positive constants. If there exists a group
of $\alpha$ and $c$ so that the interval $\Delta t/N=(t_1-t_0)/N$
vanishes in the thermodynamic limit, the percolation is said to be
explosive. In the ordinary site percolation, such $\alpha$ and $c$
never exist, leading to $\Delta t$ always of order of $N$. According
to the above definition, percolation with zero critical point can
never be explosive because the order parameter cannot have
macroscopic increase until sufficient (macroscopic) number of nodes
are occupied. Since Eq.(3) indicates that for any dense ER network
($k(N)\rightarrow \infty$ when $N\rightarrow \infty$) the critical
point is zero regardless of $m$, percolation on dense ER network
must be non-explosive and continuous. Sparsity is thus a necessary
condition for a percolation to show explosive nature. If $k$ is
finite, for $m>1$ the percolation seems immediately becoming
explosive with $\Delta t\sim o(N)$ (Fig.~\ref{explo}(a)). However
this simulation result can be deceptive and misleading, as indicated
by Ref\cite{rio}. Indeed further studies on the largest jump of the
order parameter indicates that the jump size decreases with $N$ as
power law and vanishes when $N\rightarrow
\infty$(Fig.~\ref{explo}(b)). Thus the site explosive percolation is
still continuous in the thermodynamic limit for these cases.

To find a general conclusion for all $m$, finite size scaling
analysis of $S$ for various $m$ is performed  \cite{shu,Fortu}. The
theory of finite size scaling tells us that the order parameter
obeys the relation $S=N^{-\beta/\nu}F[(t-t_c)N^{1/\nu}]$, where
$\beta$ and $\nu$ are two critical exponents and $F[*]$ is a
universal function. If the percolation is discontinuous, this
scaling relation trivially applies with $\beta=0$. Otherwise studies
of $S$ as a function of the system size $N$ yield the exponent of
$\beta$. The analysis finally gives the relation $\beta(m)\sim
m^{-1.1}$(Fig.~\ref{sca}(a)), indicating $\beta\rightarrow 0$ only
when $m\rightarrow \infty$. The finite size scaling analysis is also
applied to the susceptibility $\chi$ which obeys
$\chi=N^{\gamma/\nu}G[(t-t_c)N^{1/\nu}]$, where $\gamma$ is another
critical exponent and $G[*]$ is the related universal function. The
definition of $\chi$ along with the scaling behavior of $S$ and
$\chi$ give the relationship $\gamma/\nu+\beta/\nu=1$, which can be
used to check our measured value of the critical exponents
{\cite{Fortu}. As shown in Fig.~\ref{sca}(b), the sum
$(\beta+\gamma)/\nu$ is always $1$ with good approximation. Other
critical exponents are also reported in Fig.~\ref{sca}(b). The
power-law decay of $\beta(m)$ indicates that explosive site
percolation on ER network is continuous for any finite $m$ but could
be discontinuous when $m$ approaches infinite.

\begin{figure} \resizebox{20pc}{!}{\includegraphics{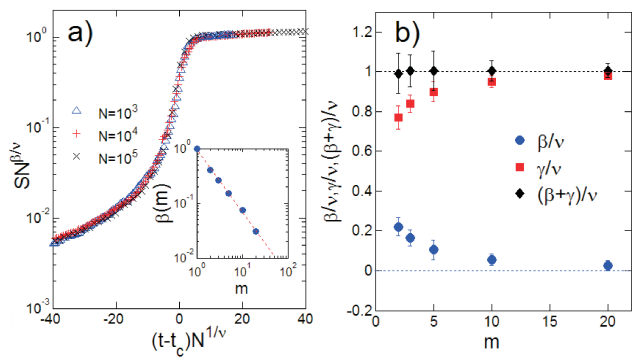}}
\vspace{-.3cm} \caption{\footnotesize  (a): Finite
size scaling analysis of the explosive percolation for the case of
$m=3$. Critical exponents can be determined by making all the data
points of different network size collapse together. For
$m=2,3,5,10,20$, $\beta\approx 0.41, 0.26, 0.15, 0.074, 0.031$ while
$\beta=1$ for $m=1$ as known for traditional site percolation.
Inset: The scaling relationship of critical exponent $\beta$ and
$m$. (b): The relationship of the critical exponents and $m$. For
$m=2,3,5,10,20$, $\beta/\nu\approx 0.221,0.164,0.105,0.055,0.025$
and $\gamma/\nu\approx 0.77,0.84,0.9,0.95,0.98$. The sum
$(\beta+\gamma)/\nu$ is always $1$ with good approximation. All the
results are obtained under network size up to $10^{5}$ nodes and are
averaged over $50$ realizations.} \label{sca}
\end{figure}

In summary, the critical points of a class of explosive site
percolation in ER random network are found to scale with the average
degree and their exponents range from $-1$ to $-\tau\approx-0.5$.
Heuristic discussions are made to uncover their possible implication
to network robustness. Note that although these results are obtained
under our specific percolation model, they could be even more
general. Since the explosive percolation, as the model proposed
here, controls directly the growth of the giant cluster, it is
expected that when $m=N$, the critical point is delayed to the
greatest extent compared to any other possible site percolation
(i.e. the upper bound provided by Eq. (3) may be valid for any site
percolation on ER network). If this assumption is true, for any site
percolation satisfying critical point $t_c>1/k$ on ER network, $t_c$
must follow Eq. (4) because such critical point is bounded by $1/k$
and Eq. (3). Thus Eq. (4) could be a general and model-independent
result. The explosive nature, according to its definition, is
deduced to occur only on sparse network. So sparsity is a necessary
condition for the explosive site percolation to be discontinuous.
When this condition is fulfilled, analysis on the critical exponents
by finite size scaling method indicates a power-law decay of
$\beta(m)$. Other critical exponents are also presented to show
their consistence. These results reveal that discontinuity of the
class of explosive site percolation could happen only in sparse
networks and only when $m\rightarrow \infty$, while for any finite
$m$ the explosive percolation is still continuous.

This work was partially supported by the National Nature Science
Foundation of China under Grant Nos. 11075057, 11035009, 10979074,
and the Shanghai Development Foundation for Science and Technology
under contract No. 09JC1416800.

{}
\end{document}